\documentclass[aps,prd,preprintnumbers,nofootinbib,superscriptaddress]{revtex4-2}

\usepackage[T1]{fontenc}
\usepackage[utf8]{inputenc}
\usepackage{lmodern}
\usepackage{amsmath,amssymb,amsfonts,mathtools,bm}
\usepackage{physics}
\usepackage{graphicx}
\usepackage{booktabs}
\usepackage{array}
\usepackage{multirow}
\usepackage{xcolor}
\usepackage{hyperref}
\usepackage[nameinlink,capitalise,noabbrev]{cleveref}
\usepackage{placeins}

\hypersetup{
  colorlinks=true,
  linkcolor=blue!60!black,
  citecolor=blue!60!black,
  urlcolor=blue!60!black
}

\providecommand{\dd}{\mathrm{d}}

\newcommand{\wre}{w_{\rm reh}}

\newcommand{\GeV}{\mathrm{GeV}}\newcommand{\MeV}{\mathrm{MeV}}
\newcommand{\Mpc}{\mathrm{Mpc}}

\begin{document}

\title{Inflation with an Exponential Power-Law Potential:\\
Analytical Slow-Roll Relations, Branch Dynamics, and Conditional Reheating Matching}

\author{Feyzollah Younesizadeh}\thanks{Corresponding author}
\email{fyounesizadeh@aut.ac.ir}
\affiliation{Department of Physics, Amirkabir University of Technology (Tehran Polytechnic), P.O. Box 15875-4413, Tehran, Iran}

\author{Younes Younesizadeh}
\email{yyounesi@aut.ac.ir}
\affiliation{Department of Physics, Amirkabir University of Technology (Tehran Polytechnic), P.O. Box 15875-4413, Tehran, Iran}

\author{Wen-Xiang Chen}
\email{wxchen4277@qq.com}
\affiliation{School of Electronic Information, Guangzhou City University of Technology, Guangzhou, China}

\date{\today}

\begin{abstract}
We study canonical single-field inflation driven by the potential
$V(\phi)=M^4\exp(\phi/M_{\rm Pl})(1+k\phi/M_{\rm Pl})^{-1/k}$ with $k<0$.
We retain the canonical Einstein--scalar action and derive analytical relations within the slow-roll approximation. Expressing the local spectra through the logarithmic slope $A=kx/(1+kx)$, with $x=\phi/M_{\rm Pl}$, yields an endpoint-independent consistency relation among the scalar tilt, tensor ratio, and running at their stated approximation orders. On the negative-field branch, an exact phase-space argument shows that an initially downhill, inflationary hilltop trajectory cannot end inflation within the bare single-field model. We specify the endpoint as a phenomenological input and use a Lambert-$W$ solution of the slow-roll e-fold integral. Tables and figures illustrate the fixed choice $x_e=-2$ and must be interpreted conditionally on a completion that implements this endpoint and preserves the adiabatic spectrum. The reheating-matching relation uses explicit scale and thermal conventions and imposes both endpoint energy bounds and nonnegative reheating duration. The observational comparison uses published marginalized intervals rather than reconstructed joint posteriors. These results establish local analytical structure and conditional numerical benchmarks. A complete exit mechanism, a unique reheating equation of state, and precision spectral predictions remain unestablished.
\end{abstract}

\maketitle


\tableofcontents
\FloatBarrier

\FloatBarrier
\section{Introduction}
\label{sec:intro}

Cosmic inflation offers a dynamical framework in which a sufficiently accelerated phase in the early Universe can address the horizon and flatness problems and dilute unwanted relics. In the same framework, quantum fluctuations of the inflaton and metric generate the primordial perturbations from which the observed large-scale structure develops \cite{Guth1981}-\cite{Bezrukov2008}. The simplest realization employs a single scalar field with a canonical kinetic term, minimally coupled to Einstein gravity. Within this setting the functional form of the scalar potential controls the slow-roll trajectory and therefore the observable scalar and tensor spectra. The theory of cosmological perturbations in such a background is well developed \cite{MukhanovChibisov1981,Mukhanov1992}, and the resulting inflationary phenomenology is reviewed in Refs.~\cite{Baumann2011,Martin2014,Odintsov2023}; for the late-time cosmological setting we refer to Ref.~\cite{DodelsonSchmidt2020}. Even with this apparent simplicity, the inflationary predictions depend sensitively on the geometry of the scalar potential and on the post-inflationary expansion history.

A broad range of single-field constructions has been explored in this framework. These include chaotic inflation and chaotic inflation in supergravity \cite{GoncharovLinde1984}, brane inflation \cite{DvaliTye1999}, hilltop inflation \cite{BoubekeurLyth2005}, natural inflation \cite{Freese1990}, the Starobinsky plateau \cite{Starobinsky1980}, Higgs inflation \cite{Bezrukov2008}, the \(\alpha\)-attractor family \cite{KalloshLinde2013,KalloshLindeRoest2014,Kallosh2013Superconformal,Carrasco2015,Linde2015}, hybrid cosmological attractors \cite{KalloshLinde2022}, and multi-field \(\alpha\)-attractors \cite{KalloshLinde2019}. String-inspired constructions such as KKLT \cite{Kachru2003} and poly-instanton inflation \cite{Cicoli2011,Kobayashi2017} provide further motivation for exponential and exponentially flattened potentials. Recent Bayesian analyses of large single-field model sets against Planck and BICEP/Keck data have refined the observational constraints and underlined the role of the reheating phase in model comparison \cite{Martin2024Crossroads,Chowdhury2019}. The question of why multi-field high-energy constructions nevertheless appear observationally as single-field has also received renewed attention \cite{Kristiano2024}. A related single-field route with a generalized exponential potential, whose shape can be smoothly deformed between plateau-like and steeper regimes, has been studied in Ref.~\cite{Younesizadeh2026GRG}.

Modified-gravity inflation offers a complementary setting in which similar inflationary behaviour can arise from the gravitational sector or from nonminimal couplings. Representative examples include \(f(R)\) gravity and its extensions \cite{Buchdahl1970}-\cite{Younesizadeh2022fRphi}, Gauss--Bonnet inflation \cite{GuoSchwarz2010}-\cite{Younesizadeh2026GB}, \(f(R,T)\) gravity \cite{Harko2011,Dzhunushaliev2014}, \(f(T)\) gravity \cite{Bengochea2009}, and trace-coupled or \(f(\phi)T\) constructions \cite{Younesizadeh2025DbraneFphiT}-\cite{Younesizadeh2026NA}. Gravitational-wave observations \cite{LIGO2016,LIGO2017} now constrain the tensor sector of such theories, notably through the propagation speed of tensor modes. These frameworks show that inflationary trajectories with flat or slowly varying regions are not restricted to the canonical Einstein--Hilbert plus scalar-field setting. They also introduce additional model dependence, generally through nonminimal couplings, higher-curvature terms, or nonconservation of the matter stress tensor. The present work remains in the canonical single-field sector so that the inflationary dynamics can be treated analytically and the model-dependent features isolated in a controlled way.

The observational benchmarks used below include the Planck 2018 constraint
\(n_s=0.9649\pm0.0042\) at \(68\%\) confidence and the BICEP/Keck bound
\(r_{0.05}<0.036\) at \(95\%\) confidence \cite{Planck2018}-\cite{Planck2018VI}. The ACT DR6 data, combined with Planck, shift the scalar tilt upward to
\(n_s=0.9709\pm0.0038\) for P-ACT. Adding ACT and Planck CMB lensing and DESI Year-1 BAO data (P-ACT-LB) gives
\(n_s=0.9743\pm0.0034\) \cite{ACTDR6,Calabrese2025}. The updated tensor constraint combining BICEP/Keck, Planck, and LIGO--Virgo--KAGRA data is
\(r<0.028\) at \(95\%\) confidence \cite{Galloni2023}. Earlier WMAP \cite{WMAP2013} and Planck 2013 \cite{Planck2013} results, the joint BICEP2/Keck/Planck analysis \cite{Ade2015}, and the BICEP/Keck 2021 analysis \cite{BICEP2021} provide further contemporary context. Future space-based interferometers such as LISA will probe inflationary stochastic gravitational-wave backgrounds beyond the vacuum tensor modes \cite{Bartolo2016LISA,LISA2024}, and the next generation of CMB polarization experiments, including LiteBIRD and CMB-S4 \cite{LiteBIRD,CMBS4}, will probe the tensor-to-scalar ratio down to the \(10^{-3}\) level. These numbers are used only for comparison; no model-specific likelihood analysis is performed here.

The post-inflationary and theoretical consistency conditions are treated in the later sections. The slow-roll matching between the pivot scale and the end of inflation requires care beyond leading order \cite{StewartLyth1993,LiddleLyth2000}. Reheating is not determined by the potential alone and must be specified through the inflaton couplings or an effective equation of state. The relevant framework includes the classic studies of reheating and preheating \cite{Kofman1994}-\cite{Allahverdi2010}, modern reviews of non-perturbative dynamics \cite{Amin2014,Barman2025}, recent analyses of plateau-type models in light of ACT DR6 \cite{Haque:2025}-\cite{Yogesh:2025}, constraints on very low reheating temperatures \cite{LowReheat2024}-\cite{Hannestad2004}, and earlier analyses of reheating-consistent inflationary parameter spaces \cite{MartinReheating2010,MartinRingevalVennin2015}. For canonical single-field inflation, the squeezed-limit non-Gaussianity obeys the consistency relation \cite{Maldacena,Lyth}, and the Planck constraints on local non-Gaussianity are given in Ref.~\cite{PlanckNG}. The effective field theory of inflation, radiative stability, and the swampland and trans-Planckian censorship conjectures provide a complementary set of consistency conditions on any ultraviolet completion \cite{EFTInflationReview}-\cite{Obied2018}. These considerations do not single out a preferred parameter region for the present model, but they constrain the class of microscopic completions and motivate the consistency analysis performed below.

The model studied here is the exponential power-law potential
\begin{equation}
V(\phi)=M^4
\exp\!\left(\frac{\phi}{M_{\rm Pl}}\right)
\left(1+k\frac{\phi}{M_{\rm Pl}}\right)^{-1/k}
,
\label{eq:intro-potential}
\end{equation}
where \(M\) is the characteristic energy scale, \(M_{\rm Pl}\) is the reduced Planck mass, and $k<0$. The logarithmic slope is elementary,
\begin{equation}
\frac{V_{,\phi}}{V}=\frac{1}{M_{\rm Pl}}\frac{k\phi/M_{\rm Pl}}{1+k\phi/M_{\rm Pl}},
\label{eq:intro-log-slope}
\end{equation}
and vanishes at the maximum $\phi=0$, where $V(0)=M^4$. The principal real domain is $x<1/|k|$. At its positive-field boundary the potential tends to zero, with the complex branch structure specified below; the potential does not diverge. Along the negative-field branch the potential decreases toward zero. Related generalized exponential single-field potentials are studied in Ref.~\cite{Younesizadeh2026GRG}; we do not assert a verified distinction from that family without a full comparison of the potential, parameters, branches, and endpoint prescriptions. Our analysis focuses on the local consistency relation, the obstruction to a bare single-field exit, and an endpoint-dependent e-fold parameterization. The numerical benchmarks and their observational interpretation are collected in Sec.~\ref{sec:closedform-scan}.

The paper is organized as follows. Section~\ref{sec:framework} introduces the Einstein-scalar system. Section~\ref{sec:geometry} analyzes the geometry and limiting behavior of the potential. Section~\ref{sec:slowroll} derives the slow-roll parameters and the inflationary e-fold relation. Section~\ref{sec:perturbations} presents the scalar and tensor perturbations, and Section~\ref{sec:eft} discusses the field excursion and effective-field-theory diagnostics. Section~\ref{sec:reheating} treats the exit and reheating problem. Section~\ref{sec:observational} formulates the observational analysis and numerical strategy, and Section~\ref{sec:closedform-scan} records the closed-form algebraic parameterization with its numerical scan and the associated figures, which together constitute the primary result of the paper. Section~\ref{sec:comparison} compares the present construction with representative inflationary classes. Section~\ref{sec:discussion} contains a discussion, and we conclude in Sec.~\ref{sec:conclusions}.

\FloatBarrier
\section{Canonical single-field inflation}
\label{sec:framework}

\subsection{Action and background equations}

We consider Einstein gravity minimally coupled to a canonical scalar,
\begin{equation}
S=\int\dd^4x\,\sqrt{-g}
\left[
\frac{M_{\rm Pl}^2}{2}R
-\frac12 g^{\mu\nu}\partial_\mu\phi\partial_\nu\phi
-V(\phi)
\right].
\label{eq:action}
\end{equation}
The reduced Planck mass is
\begin{equation}
M_{\rm Pl}=\frac{1}{\sqrt{8\pi G}}.
\end{equation}

For a spatially flat FLRW geometry,
\begin{equation}
\dd s^2=-\dd t^2+a^2(t)\dd{\bm x}^2,
\label{eq:flrw}
\end{equation}
the energy density and pressure of a homogeneous scalar are
\begin{align}
\rho_\phi&=\frac12\dot\phi^2+V(\phi),\\
p_\phi&=\frac12\dot\phi^2-V(\phi).
\end{align}
The Friedmann equations are
\begin{align}
H^2&=\frac{1}{3M_{\rm Pl}^2}
\left(\frac12\dot\phi^2+V\right),
\label{eq:friedmann}\\
\dot H&=-\frac{\dot\phi^2}{2M_{\rm Pl}^2},
\label{eq:hdot}
\end{align}
while the Klein--Gordon equation is
\begin{equation}
\ddot\phi+3H\dot\phi+V_{,\phi}=0.
\label{eq:kg}
\end{equation}

The acceleration equation reads
\begin{equation}
\frac{\ddot a}{a}
=-\frac{1}{6M_{\rm Pl}^2}
\left(\rho_\phi+3p_\phi\right)
=\frac{1}{3M_{\rm Pl}^2}
\left(V-\dot\phi^2\right).
\end{equation}
Hence inflation occurs when
\begin{equation}
\dot\phi^2<V.
\end{equation}

\subsection{Hubble slow-roll parameter}

The exact Hubble slow-roll parameter is
\begin{equation}
\epsilon_H\equiv-\frac{\dot H}{H^2}
=\frac{\dot\phi^2}{2M_{\rm Pl}^2H^2}.
\label{eq:epsilonH}
\end{equation}
Inflation corresponds to
\begin{equation}
\epsilon_H<1.
\end{equation}
The exact relation
\begin{equation}
\frac{\ddot a}{a}=H^2(1-\epsilon_H)
\end{equation}
makes the physical meaning of $\epsilon_H$ transparent.

\subsection{Potential slow roll}

When
\begin{equation}
\dot\phi^2\ll V,
\qquad
|\ddot\phi|\ll3H|\dot\phi|,
\end{equation}
the background equations reduce to
\begin{align}
H^2&\simeq\frac{V}{3M_{\rm Pl}^2},
\label{eq:srH}\\
3H\dot\phi&\simeq-V_{,\phi}.
\label{eq:srKG}
\end{align}
The potential slow-roll parameters are
\begin{align}
\epsilon_V&\equiv
\frac{M_{\rm Pl}^2}{2}
\left(\frac{V_{,\phi}}{V}\right)^2,
\label{eq:epsV}\\
\eta_V&\equiv
M_{\rm Pl}^2\frac{V_{,\phi\phi}}{V}.
\label{eq:etaV}
\end{align}
To first order,
\begin{equation}
\epsilon_H\simeq\epsilon_V,
\end{equation}
and inflation requires $\epsilon_V<1$ in the slow-roll regime.

\FloatBarrier
\section{Geometry of the exponential power-law potential}
\label{sec:geometry}

\subsection{Dimensionless field variable}

Introduce
\begin{equation}
x\equiv\frac{\phi}{M_{\rm Pl}}.
\label{eq:xdef}
\end{equation}
Then Eq.~\eqref{eq:intro-potential} becomes
\begin{equation}
V(x)=M^4e^x(1+kx)^{-1/k}.
\label{eq:Vx}
\end{equation}
The potential is real for the principal real branch provided
\begin{equation}
1+kx>0.
\end{equation}
For $k<0$ this requires $x<1/|k|$, so the negative field values relevant to the present analysis are automatically within the domain. Write $p=-1/k>0$ and $x_b=-1/k$. As $x\to x_b^-$, $V\propto(1+kx)^p\to0$. Under complex continuation, $x_b$ is generally a branch point for noninteger $p$ and an analytic zero for positive integer $p$. The logarithmic derivative, rather than the potential itself, has a simple pole there. No spacetime singularity follows from this field-space boundary.

It is useful to define
\begin{equation}
\Delta\equiv1+kx.
\label{eq:Deltadef}
\end{equation}
Then
\begin{equation}
x=\frac{\Delta-1}{k}.
\end{equation}

\subsection{Logarithmic derivatives}

The logarithm of the potential is
\begin{equation}
\ln V
=
4\ln M+x-\frac1k\ln(1+kx).
\end{equation}
Therefore,
\begin{equation}
\frac{1}{V}\frac{\dd V}{\dd x}
=
1-\frac{1}{1+kx}
=
\frac{kx}{1+kx}.
\label{eq:dlnVdx}
\end{equation}
Since $\dd x/\dd\phi=M_{\rm Pl}^{-1}$,
\begin{equation}
\frac{V_{,\phi}}{V}
=
\frac{1}{M_{\rm Pl}}
\frac{kx}{1+kx}.
\label{eq:VprimeoverV}
\end{equation}

Differentiating once more,
\begin{equation}
\frac{\dd}{\dd x}
\left(\frac{kx}{1+kx}\right)
=
\frac{k}{(1+kx)^2},
\end{equation}
and hence
\begin{equation}
\frac{V_{,\phi\phi}}{V}
=
\frac{1}{M_{\rm Pl}^2}
\left[
\frac{k}{(1+kx)^2}
+
\left(\frac{kx}{1+kx}\right)^2
\right].
\label{eq:V2overV}
\end{equation}

\subsection{Stationary point and curvature}

Equation~\eqref{eq:VprimeoverV} shows that
\begin{equation}
V_{,\phi}=0
\quad\Longleftrightarrow\quad
x=0.
\end{equation}
At this point,
\begin{equation}
V(0)=M^4,
\end{equation}
and
\begin{equation}
V_{,\phi\phi}(0)=\frac{kM^4}{M_{\rm Pl}^2}.
\end{equation}
Since $k<0$, the curvature is negative and $x=0$ is a maximum. The potential has the profile of a broad hilltop, with the field driven away from $x=0$ along the negative-$x$ direction and toward $x\to-\infty$, where $V$ asymptotes to zero.

Expanding around $x=0$ gives
\begin{equation}
V(\phi)
=
M^4
\left[
1+\frac{k}{2}x^2
-\frac{k^2}{3}x^3
+\mathcal{O}(x^4)
\right],
\label{eq:Vsmallx}
\end{equation}
so that
\begin{equation}
V(\phi)
\simeq
M^4+\frac12m_\phi^2\phi^2,
\end{equation}
with
\begin{equation}
m_\phi^2=\frac{kM^4}{M_{\rm Pl}^2}.
\label{eq:mphi}
\end{equation}
For $k<0$, $m_\phi^2<0$, the stationary point is unstable, and no oscillatory phase can occur about $x=0$.

There is no local minimum for $k<0$. The exact phase-space argument in Sec.~\ref{sec:reheating} establishes the stronger dynamical obstruction to ending inflation on the downhill negative-field branch. The absence of a minimum alone would not prove this obstruction.

\subsection{Asymptotic limits}

For $k\rightarrow0$ at fixed $x$,
\begin{equation}
\ln(1+kx)
=
kx-\frac{k^2x^2}{2}+\mathcal{O}(k^3),
\end{equation}
and therefore
\begin{equation}
\ln\frac{V}{M^4}
=
x-\frac1k\ln(1+kx)
=
\frac{kx^2}{2}+\mathcal{O}(k^2).
\end{equation}
Thus
\begin{equation}
k\rightarrow0:\qquad V\rightarrow M^4.
\label{eq:klow}
\end{equation}
The $k\rightarrow0$ limit is a vacuum-energy limit.

For fixed positive $x$ and $k\rightarrow\infty$,
\begin{equation}
\frac1k\ln(1+kx)\rightarrow0,
\end{equation}
and hence
\begin{equation}
k\rightarrow\infty:\qquad V\rightarrow M^4e^x.
\label{eq:khigh}
\end{equation}
The large-$k$ limit is exponential rather than cosmological-constant-like.

For $k<0$ and $x\to-\infty$ the exponential factor dominates the power-law factor. A short calculation on the logarithm,
\begin{equation}
\ln\frac{V}{M^4}
=
x-\frac1k\ln(1+kx)
\sim
x+\frac{1}{|k|}\ln(|k||x|),
\end{equation}
shows that $V/M^4\to0^+$ as $x\to-\infty$. Since $\dd\ln V/\dd x\to1$, this region is an exponential tail rather than a plateau with a vanishing relative slope. The right flank reaches the real-domain boundary $x=1/|k|$, where the potential tends to zero.

\FloatBarrier
\section{Slow-roll dynamics}
\label{sec:slowroll}

\subsection{Slow-roll parameters}

Substitution of Eq.~\eqref{eq:VprimeoverV} into Eq.~\eqref{eq:epsV} gives
\begin{equation}
\epsilon_V
=
\frac12
\left(
\frac{kx}{1+kx}
\right)^2
=
\frac12
\left(
1-\frac1\Delta
\right)^2.
\label{eq:epsVexplicit}
\end{equation}
Similarly,
\begin{equation}
\eta_V
=
\frac{k}{(1+kx)^2}
+
\left(
\frac{kx}{1+kx}
\right)^2
=
\frac{k}{\Delta^2}
+
\left(1-\frac1\Delta\right)^2.
\label{eq:etaVexplicit}
\end{equation}

A useful third potential slow-roll parameter is
\begin{equation}
\xi_V^2
\equiv
M_{\rm Pl}^4
\frac{V_{,\phi}V_{,\phi\phi\phi}}{V^2}.
\label{eq:xivdef}
\end{equation}
Writing
\begin{equation}
A(x)\equiv\frac{\dd\ln V}{\dd x}=\frac{kx}{1+kx},
\qquad \frac{\dd A}{\dd x}=k(1-A)^2,
\label{eq:Astructure}
\end{equation}
we have
\begin{equation}
\frac{V_{,\phi}}V=\frac{A}{M_{\rm Pl}},
\end{equation}
and
\begin{equation}
M_{\rm Pl}^2\frac{V_{,\phi\phi}}V
=A^2+\frac{k}{(1+kx)^2}.
\end{equation}
A further differentiation yields
\begin{equation}
M_{\rm Pl}^3\frac{V_{,\phi\phi\phi}}V
=
A^3+\frac{3kA}{(1+kx)^2}
-\frac{2k^2}{(1+kx)^3}.
\end{equation}
Hence
\begin{equation}
\xi_V^2
=
A
\left[
A^3+\frac{3kA}{\Delta^2}
-\frac{2k^2}{\Delta^3}
\right].
\label{eq:xiv}
\end{equation}

\subsection{End of slow roll}

The formal condition $\epsilon_V=1$ gives
\begin{equation}
\frac12
\left(\frac{kx_e}{1+kx_e}\right)^2=1,
\end{equation}
or
\begin{equation}
\left|\frac{kx_e}{1+kx_e}\right|=\sqrt2.
\end{equation}
The two algebraic roots are
\begin{equation}
x_e^{(+)}
=
\frac{\sqrt2}{k(1-\sqrt2)},
\qquad
x_e^{(-)}
=
-\frac{\sqrt2}{k(1+\sqrt2)}.
\end{equation}
For $k<0$, the admissible root is
\begin{equation}
x_{\epsilon_V=1}=\frac{2-\sqrt2}{|k|}>0,
\end{equation}
on the positive-field branch; the other root lies outside $1+kx>0$.
For example, $k=-0.016$ gives $x_{\epsilon_V=1}=36.61165$, below the boundary $62.5$. On the negative-field branch, $0<A<1$ and $0<\epsilon_V<1/2$, so this positive-field root cannot be used as its endpoint. We therefore specify $x_e<0$ as a matching input. The exact no-exit statement below does not rely on identifying $\epsilon_H$ with $\epsilon_V$.

\subsection{Number of e-folds}

Let $N=\ln(a_e/a_*)$ denote the number of e-folds remaining to the same specified endpoint, with $x_e<x_*<0$. The slow-roll number of e-folds is
\begin{equation}
N
\simeq
\frac{1}{M_{\rm Pl}^2}
\int_{\phi_{\rm end}}^{\phi_*}
\frac{V}{V_{,\phi}}\dd\phi.
\label{eq:Ngeneral}
\end{equation}
Using Eq.~\eqref{eq:VprimeoverV},
\begin{equation}
\frac{V}{V_{,\phi}}
=
M_{\rm Pl}\frac{1+kx}{kx},
\end{equation}
and $\dd\phi=M_{\rm Pl}\dd x$, we obtain
\begin{equation}
N
=
\int_{x_e}^{x_*}
\frac{1+kx}{kx}\dd x.
\end{equation}
Therefore,
\begin{equation}
N
=
(x_*-x_e)
+
\frac1k
\ln\left|\frac{x_*}{x_e}\right|.
\label{eq:Nx}
\end{equation}
In terms of $\Delta$,
\begin{equation}
N
=
\frac1k
\left[
\Delta_*-\Delta_e
+
\ln\left|
\frac{\Delta_*-1}{\Delta_e-1}
\right|
\right].
\label{eq:ND}
\end{equation}

Equations~\eqref{eq:Nx} and~\eqref{eq:ND} are integrated slow-roll relations, not exact solutions of the full background equations. For fixed $x_e<0$ their real inverse is
\begin{equation}
x_*(N;k,x_e)=\frac1k W_0\!\left(kx_e\,e^{kx_e+kN}\right),
\label{eq:Lambert}
\end{equation}
where the argument is positive and $W_0$ is the principal real Lambert branch. Differentiation gives $\dd x_*/\dd N=A(x_*)>0$, consistent with Eq.~\eqref{eq:dxN}. If an auxiliary sector triggers the exit, $x_e$ is an independent input until that sector is specified. The numerical example fixes $x_e=-2$ at every scan point; this choice does not supply an exit mechanism.

\subsection{Field velocity}

The slow-roll equation gives
\begin{equation}
\dot\phi\simeq-\frac{V_{,\phi}}{3H},
\end{equation}
and hence
\begin{equation}
\frac{\dd\phi}{\dd N}
\simeq
M_{\rm Pl}\sqrt{2\epsilon_V}
\end{equation}
on the negative-field branch, where $N$ is the remaining e-fold count, $\dd N/\dd t=-H$, and $V_{,\phi}>0$. In terms of $x$,
\begin{equation}
\frac{\dd x}{\dd N}
\simeq
\frac{kx}{1+kx}.
\label{eq:dxN}
\end{equation}

\FloatBarrier
\section{Primordial perturbations}
\label{sec:perturbations}

\subsection{Scalar power spectrum}

For a canonical single-field model with Bunch--Davies initial conditions, the leading slow-roll scalar spectrum at horizon crossing is
\begin{equation}
\mathcal P_{\cal R}(q)
\simeq
\left.
\frac{H^2}{8\pi^2M_{\rm Pl}^2\epsilon_H}
\right|_{q=aH}.
\end{equation}
Using the potential approximation,
\begin{equation}
A_s
\simeq
\frac{V_*}
{24\pi^2M_{\rm Pl}^4\epsilon_{V*}}.
\label{eq:As}
\end{equation}
For the present potential,
\begin{equation}
A_s
=
\frac{M^4}{24\pi^2M_{\rm Pl}^4\epsilon_{V*}}
e^{x_*}
(1+kx_*)^{-1/k}.
\end{equation}
Thus the normalization fixes
\begin{equation}
\frac{M}{M_{\rm Pl}}
=
\left[
24\pi^2A_s\epsilon_{V*}
e^{-x_*}
(1+kx_*)^{1/k}
\right]^{1/4}.
\label{eq:Mnormalization}
\end{equation}

\subsection{Scalar spectral index}

At first order in slow roll,
\begin{equation}
n_s-1
=
-6\epsilon_V+2\eta_V.
\end{equation}
For the exponential power-law potential,
\begin{equation}
n_s-1
=
\frac{2k}{(1+kx_*)^2}
-
\left(
\frac{kx_*}{1+kx_*}
\right)^2.
\label{eq:ns}
\end{equation}
Equivalently,
\begin{equation}
n_s
=
1+\frac{2k}{\Delta_*^2}
-\left(1-\frac1{\Delta_*}\right)^2.
\end{equation}

\subsection{Tensor spectrum}

The tensor spectrum is
\begin{equation}
\mathcal P_T
\simeq
\frac{2H_*^2}{\pi^2M_{\rm Pl}^2}.
\end{equation}
The tensor-to-scalar ratio is
\begin{equation}
r=16\epsilon_{V*}
=
8\left(
\frac{kx_*}{1+kx_*}
\right)^2.
\label{eq:r}
\end{equation}
The tensor spectral index is
\begin{equation}
n_t=-2\epsilon_{V*}
=
-\left(
\frac{kx_*}{1+kx_*}
\right)^2,
\label{eq:nt}
\end{equation}
and the standard single-field consistency relation is
\begin{equation}
r=-8n_t.
\label{eq:consistency}
\end{equation}

\subsection{Running of the scalar spectral index}

Define
\begin{equation}
\alpha_s\equiv\frac{\dd n_s}{\dd\ln q}.
\end{equation}
At leading nontrivial order,
\begin{equation}
\alpha_s
=
16\epsilon_V\eta_V
-24\epsilon_V^2
-2\xi_V^2.
\label{eq:alphasgeneral}
\end{equation}
Substitution of Eqs.~\eqref{eq:epsVexplicit}, \eqref{eq:etaVexplicit}, and \eqref{eq:xiv} yields
\begin{equation}
\alpha_s
=
\frac{2k^3x_*(x_*+2)}{(1+kx_*)^4}
\label{eq:alphas}
\end{equation}
where $q$ is the comoving wavenumber and $\dd\ln q\simeq-\dd N$. An independent chain-rule check follows because $-(\dd n_s/\dd x)A$ reproduces Eq.~\eqref{eq:alphas}. For precision work, the running should be evaluated from the full slow-roll hierarchy or exact numerical mode evolution.

\subsection{Local consistency relation and the sign of the running}

Using $A$, the first-order spectra read
\begin{equation}
r=8A^2,\qquad n_s-1=-A^2+2k(1-A)^2.
\end{equation}
Define $u=\sqrt{r/8}=A$ and $\delta=1-n_s$. Eliminating $x$ and $k$ gives
\begin{equation}
k=\frac{u^2-\delta}{2(1-u)^2},\qquad
\alpha_s=\frac{u(u^2-\delta)(u-\delta)}{1-u}.
\label{eq:local-consistency}
\end{equation}
This relation uses first-order $n_s,r$ and the first nonzero slow-roll order of $\alpha_s$; it is not an all-order identity for exact spectra. It is independent of the endpoint and of the e-fold inversion. On the negative-field branch $k<0$ requires $\delta>u^2$, and
\begin{equation}
\alpha_s>0\ \Longleftrightarrow\ u<\delta
\ \Longleftrightarrow\ r<8(1-n_s)^2
\ \Longleftrightarrow\ -2<x_*<0.
\label{eq:running-sign}
\end{equation}
The running vanishes at $x_*=-2$ and becomes negative farther down the branch. Positive running is therefore confined to a restricted window. Equation~\eqref{eq:local-consistency} provides an internal falsification test at the stated orders; its uniqueness relative to related potentials and its practical detectability are not established here. Using this small running as a precision discriminator requires a second-order and exact-mode error estimate.

\subsection{Inflationary energy scale}

The scalar amplitude and tensor ratio imply
\begin{equation}
V_*=
\frac{3\pi^2}{2}A_srM_{\rm Pl}^4,
\label{eq:Vstar}
\end{equation}
and therefore
\begin{equation}
H_*=
\pi M_{\rm Pl}
\sqrt{\frac{A_sr}{2}}.
\label{eq:Hstar}
\end{equation}
The corresponding characteristic inflationary energy scale is
\begin{equation}
V_*^{1/4}
=
\left(
\frac{3\pi^2}{2}A_sr
\right)^{1/4}
M_{\rm Pl}.
\label{eq:Vquarter}
\end{equation}

\FloatBarrier
\section{Field excursion and effective-field-theory diagnostics}
\label{sec:eft}

The inflaton field excursion satisfies the Lyth relation
\begin{equation}
\frac{\dd\phi}{\dd N}
\simeq
M_{\rm Pl}\sqrt{\frac{r}{8}},
\end{equation}
so that
\begin{equation}
\frac{\Delta\phi}{M_{\rm Pl}}
\simeq
\int_{N_{\rm end}}^{N_*}
\sqrt{\frac{r(N)}{8}}\,\dd N.
\label{eq:lyth}
\end{equation}
For approximately constant $r$ this reduces to
\begin{equation}
\frac{\Delta\phi}{M_{\rm Pl}}
\sim
N_*\sqrt{\frac{r}{8}}.
\end{equation}

The dimensionless potential gradients are
\begin{align}
M_{\rm Pl}\frac{|V_{,\phi}|}{V}
&=
\left|
\frac{kx}{1+kx}
\right|,
\label{eq:EFTgradient}\\
M_{\rm Pl}^2\frac{V_{,\phi\phi}}{V}
&=
\frac{k}{(1+kx)^2}
+
\left(
\frac{kx}{1+kx}
\right)^2.
\label{eq:EFTcurvature}
\end{align}
These quantities can be used as diagnostics of the flatness and field-range assumptions of an effective description. They do not, by themselves, establish or exclude a particular ultraviolet completion.

The potential can be expanded for $|kx|<1$ as
\begin{equation}
V
=
M^4
\exp\left[
\frac{kx^2}{2}
-\frac{k^2x^3}{3}
+\frac{k^3x^4}{4}
+\mathcal{O}(k^4x^5)
\right].
\label{eq:EFTexpansion}
\end{equation}
This makes explicit that the $k\rightarrow0$ theory approaches a constant potential.

\FloatBarrier
\section{Exit from inflation and the reheating problem}
\label{sec:reheating}

\subsection{Why the bare potential is incomplete}

For $k<0$ the field-space geometry has an immediate dynamical implication. Since
\begin{equation}
V_{,\phi}\gtrless0
\quad\text{for}\quad
\phi\lessgtr0,
\end{equation}
the force $-V_{,\phi}$ points away from $\phi=0$ on both sides, so $\phi=0$ is a repeller of the homogeneous classical dynamics and the field is driven down the left flank toward negative $\phi$.

At the stationary point,
\begin{equation}
V_{\max}=M^4.
\end{equation}
The potential has no minimum for $k<0$: it decreases along the negative-$x$ branch and tends to zero as $x\to-\infty$. The exact dynamics establishes a stronger statement.

\noindent\textbf{Proposition (no downhill exit in the bare model).}
In an expanding, spatially flat universe containing only this canonical scalar with $V>0$, a solution starting at $x<0$ and $0\le v<1/\sqrt3$, where
\begin{equation}
v=-\frac{\dot\phi}{\sqrt6 M_{\rm Pl}H},\qquad n=\ln a,
\end{equation}
remains inflationary and cannot cross to the positive-field branch. To see this, the exact Einstein--scalar equations imply
\begin{equation}
\frac{\dd v}{\dd n}=(1-v^2)\left[\sqrt{\frac32}A-3v\right],
\qquad\frac{\dd x}{\dd n}=-\sqrt6 v,\qquad\epsilon_H=3v^2.
\label{eq:exact-phase}
\end{equation}
For $x<0$, $0<A<1$. The vector field points inward at $v=0$ and at $v=1/\sqrt3$: at the latter boundary it is $(2/3)[\sqrt{3/2}A-\sqrt3]<0$. The strip is therefore forward invariant, while $\dd x/\dd n\le0$. The statement follows under the specified assumptions. It does not cover initially kinetic-dominated states, extra fields, a modified potential, or energy transfer.

A complete cosmological history requires an additional ingredient to end inflation and reheat the universe. Several routes are possible. A hybrid transition may terminate inflation at a field value $\phi_{\rm exit}$ before the asymptotic regime is reached. A coupling to additional fields may modify the effective potential so that a minimum develops. A nonperturbative transition may carry the field to a different vacuum. Each route changes the relation between the inflationary sector and the post-inflationary history, and the choice must be made explicit in any complete model. The following discussion is conditional on such a completion. Figure~\ref{fig:reheating-param} tests the necessary matching conditions without constructing the completion.

\subsection{Effective reheating equation of state}

For an approximately constant average reheating equation of state $\wre$, the energy density scales as
\begin{equation}
\rho(a)\propto a^{-3(1+\wre)}.
\end{equation}
Therefore,
\begin{equation}
\rho_{\rm reh}
=
\rho_{\rm end}
\exp[-3(1+\wre)N_{\rm reh}].
\label{eq:rho_re}
\end{equation}
Thus
\begin{equation}
N_{\rm reh}
=
\frac{1}{3(1+\wre)}
\ln\left(
\frac{\rho_{\rm end}}{\rho_{\rm reh}}
\right).
\label{eq:Nre}
\end{equation}

At thermalization,
\begin{equation}
\rho_{\rm reh}
=
\frac{\pi^2}{30}g_*(T_{\rm reh})T_{\rm reh}^4,
\end{equation}
so
\begin{equation}
N_{\rm reh}
=
\frac{1}{3(1+\wre)}
\ln\left[
\frac{30\rho_{\rm end}}
{\pi^2g_*T_{\rm reh}^4}
\right].
\label{eq:NreT}
\end{equation}

\subsection{Perturbative reheating in a completed model}

Suppose a completed potential has a quadratic minimum,
\begin{equation}
V_{\rm c}(\phi)
=
\frac12m_\phi^2\phi^2+\cdots.
\end{equation}
A Yukawa interaction
\begin{equation}
\mathcal L_{\rm int}
=
-y\phi\bar\psi\psi
\end{equation}
gives, for sufficiently light fermions,
\begin{equation}
\Gamma_{\phi\rightarrow\bar\psi\psi}
\simeq
\frac{y^2m_\phi}{8\pi}.
\label{eq:GammaYukawa}
\end{equation}
A scalar interaction
\begin{equation}
\mathcal L_{\rm int}
=
-\frac{\mu}{2}\phi\chi^2
\end{equation}
similarly gives
\begin{equation}
\Gamma_{\phi\rightarrow\chi\chi}
\simeq
\frac{\mu^2}{32\pi m_\phi}
\end{equation}
when kinematically allowed.

The reheating temperature is estimated by setting $H\simeq\Gamma_\phi$,
\begin{equation}
T_{\rm reh}
\simeq
\left(
\frac{90}{\pi^2g_*(T_{\rm reh})}
\right)^{1/4}
\sqrt{\Gamma_\phi M_{\rm Pl}}.
\label{eq:Treh}
\end{equation}
A radiation era must begin well before primordial nucleosynthesis, requiring a reheating temperature above the MeV scale, although the exact lower bound depends on the detailed thermal history.

\subsection{Pivot-scale e-folds}

We denote the pivot wavenumber by $q_*=a_*H_*$ to distinguish it from the potential parameter $k$. Scale matching and entropy conservation after thermalization give \cite{Cook2015}
\begin{align}
N_*&=-\ln\frac{q_*}{a_0H_0}-N_{\rm reh}
+\ln\frac{a_{\rm reh}}{a_0}+\ln\frac{H_*}{H_0},
\label{eq:scale-matching}\\
\frac{a_{\rm reh}}{a_0}&=\frac{T_0}{T_{\rm reh}}
\left(\frac{g_{s0}}{g_{s,\rm reh}}\right)^{1/3}.
\end{align}
Substituting Eq.~\eqref{eq:Nre}, and using $H_*^2\simeq V_*/(3M_{\rm Pl}^2)$, gives
\begin{align}
N_*\simeq{}&C-\ln\frac{q_*}{a_0H_0}
+\frac14\ln\frac{V_*}{M_{\rm Pl}^4}
+\frac14\ln\frac{V_*}{\rho_{\rm end}}\nonumber\\
&+\frac{1-3\wre}{12(1+\wre)}
\ln\frac{\rho_{\rm reh}}{\rho_{\rm end}},
\label{eq:Nstarreheating}
\end{align}
with
\begin{equation}
C=\ln\frac{T_0}{H_0}-\frac12\ln3
+\frac14\ln\frac{\pi^2g_*}{30}
+\frac13\ln\frac{g_{s0}}{g_{s,\rm reh}}.
\end{equation}
All energies use the same units. The final logarithmic contribution has a positive coefficient as written; a numerical additive constant must use consistent pivot and thermal conventions. For numerical evaluation, we set $a_0=1$ and use the equivalent form without $H_0$,
\begin{align}
L(T)&=-N_*+\ln\frac{T_0}{T}
+\frac13\ln\frac{g_{s0}}{g_{s,\rm reh}}+\ln\frac{H_*}{q_*},\\
\rho_{\rm end}^{\rm match}(T,w)&=\rho_{\rm reh}(T)
\exp[3(1+w)L(T)],\qquad N_{\rm reh}=L(T).
\label{eq:rho-match}
\end{align}
At fixed $N_*,H_*$ and degrees of freedom,
\begin{equation}
\frac{\dd\ln\rho_{\rm end}^{\rm match}}{\dd\ln T}=1-3w,
\label{eq:reheat-slope}
\end{equation}
so the slopes for $w=0,1/3,1$ are $+1,0,-2$.
For $w>-1$ and no additional energy injection, the necessary conditions are
\begin{equation}
\rho_{\rm reh}\le\rho_{\rm end}\le\rho_*,\qquad N_{\rm reh}\ge0.
\label{eq:reheat-bounds}
\end{equation}
Here $\rho_*$ and $\rho_{\rm end}$ are total energy densities, including any exit sector; replacing $\rho_*$ by $V_*$ is a slow-roll approximation. A specified completion must also relate $\rho_{\rm end}$ to the energy at $x_e$; both quantities cannot be treated as freely adjustable predictions. The figure below is a conditional matching diagnostic, not proof of such a realization.

\FloatBarrier
\section{Observational constraints and numerical strategy}
\label{sec:observational}

\subsection{Observational targets}

The Planck 2018 inflation analysis finds
\begin{equation}
n_s=0.9649\pm0.0042
\end{equation}
at $68\%$ confidence in the baseline analysis \cite{Planck2018}. The scalar amplitude at the conventional pivot scale is of order
\begin{equation}
A_s\simeq2.1\times10^{-9}.
\end{equation}
Tensor constraints from the Planck and BICEP/Keck combination are at the few-percent level \cite{BICEP2021}. These numbers should be regarded as observational benchmarks rather than as a substitute for a likelihood analysis.

\subsection{Parameter-space map}

A numerical analysis should scan
\begin{equation}
(k,x_e,N_*,\wre,T_{\rm reh})
\end{equation}
or an equivalent set of independent parameters. For each point, one first solves the e-fold relation for the horizon-exit field $x_*$ and verifies the physical domain $1+kx_*>0$. One then evaluates $\epsilon_{V*}$ and $\eta_{V*}$, computes $n_s$, $r$, and $\alpha_s$, and uses $A_s$ to determine $M$. The next step is to calculate $H_*$ and $V_*^{1/4}$, as well as the field excursion. After that, $\rho_{\rm end}$ is specified through the chosen exit mechanism, the reheating relation is imposed, and points that fail the effective-field-theory or cosmological consistency conditions are discarded.

The principal diagnostic plane is
\begin{equation}
(n_s,r).
\end{equation}
Additional useful projections include
\begin{align}
n_s&=n_s(k,N_*;x_e),\\
r&=r(k,N_*;x_e),\\
\alpha_s&=\alpha_s(k,N_*;x_e),\\
T_{\rm reh}&=T_{\rm reh}(k,N_*,\wre;\text{exit inputs}),\\
N_{\rm reh}&=N_{\rm reh}(k,N_*,\wre;\text{exit inputs}).
\end{align}

\subsection{Recommended numerical implementation}

The exact background equations should also be integrated numerically. A convenient dimensionless time is
\begin{equation}
\tau=Mt.
\end{equation}
Writing
\begin{equation}
x=\frac{\phi}{M_{\rm Pl}},
\end{equation}
the Klein--Gordon equation becomes
\begin{equation}
\frac{\dd^2x}{\dd\tau^2}
+
3\mathcal H\frac{\dd x}{\dd\tau}
+
\frac{1}{M^2M_{\rm Pl}^2}
\frac{\dd V}{\dd x}
=0,
\end{equation}
where
\begin{equation}
\mathcal H\equiv\frac{H}{M}.
\end{equation}
The Friedmann constraint is
\begin{equation}
\mathcal H^2
=
\frac{M_{\rm Pl}^2}{3M_{\rm Pl}^2}
\left[
\frac12\left(\frac{\dd x}{\dd\tau}\right)^2
+\frac{V}{M^4}
\frac{M^2}{M_{\rm Pl}^2}
\right],
\end{equation}
with the normalization chosen consistently with the numerical implementation.

In practice it is simpler to integrate the original equations in dimensionless variables after fixing $M/M_{\rm Pl}$ from $A_s$. The numerical evolution should monitor
\begin{equation}
\epsilon_H=-\frac{\dot H}{H^2}
\end{equation}
and compare it against $\epsilon_V$.

\FloatBarrier
\section{Closed-form results for the $k<0$ regime}
\label{sec:closedform-scan}

All results below use the same specified endpoint, $x_e=-2$, for every value of $k$ and every e-fold label. This is a phenomenological matching choice, not a prediction of the bare potential. The labels $N=50,55,60$ denote remaining e-folds evaluated in slow roll; exact background durations differ by the truncation errors reported below. The local spectra remain conditional on a completion that implements this endpoint without changing the relevant adiabatic perturbations.

\begin{figure*}[htbp]
\centering
\includegraphics[width=\textwidth]{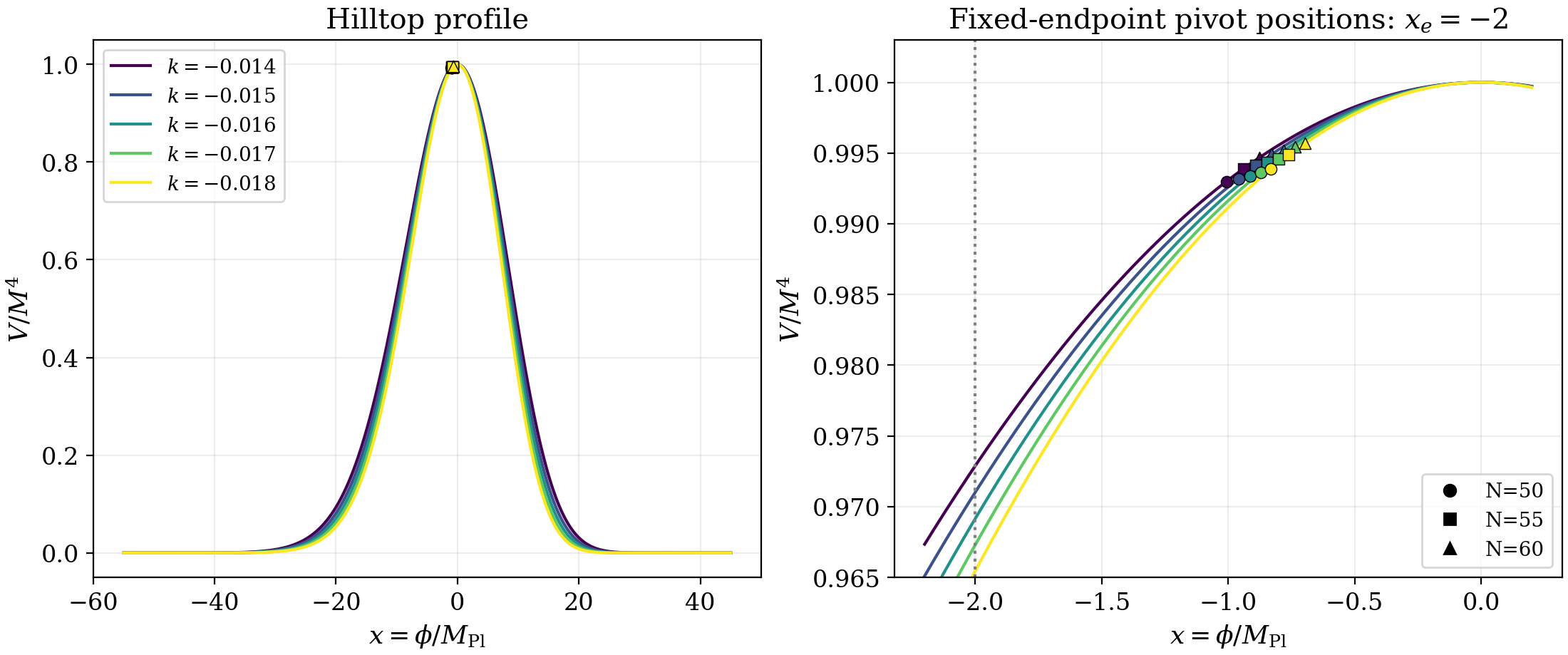}
\caption{Potential profile $V/M^4$ for the five scanned values of $k$. The right panel resolves the pivot window; circles, squares, and triangles mark $N=50,55,60$, respectively, using Eq.~\eqref{eq:compact-N} with the same prescribed endpoint $x_e=-2$. All pivots are near the hilltop. The positive-field real-domain boundary lies outside the plotted range and is not a divergent pole of $V$.}
\label{fig:potential-kneg}
\end{figure*}

The field--e-fold relation and its inverse are
\begin{equation}
N=(x_*-x_e)+\frac1k\ln\frac{x_*}{x_e},\qquad
X(N;k,x_e)=\frac1k W_0\!\left(kx_e e^{kx_e+kN}\right)=x_*.
\label{eq:compact-N}
\end{equation}
With $X=X(N;k,x_e)$, the local observables become
\begin{align}
r(N,k;x_e)&=8\left(\frac{kX}{1+kX}\right)^2,
\label{eq:compact-r}\\
n_s(N,k;x_e)&=1+\frac{2k}{(1+kX)^2}
-\left(\frac{kX}{1+kX}\right)^2,
\label{eq:compact-ns}\\
\alpha_s(N,k;x_e)&=\frac{2k^3X(X+2)}{(1+kX)^4}.
\label{eq:compact-running}
\end{align}
These expressions are evaluated without intermediate rounding. The convention is $\alpha_s=\dd n_s/\dd\ln q$, with $\dd\ln q\simeq-\dd N$. The Lambert inversion, the flow $\dd x_*/\dd N=A>0$, and the chain-rule running are mutually consistent. Figure~\ref{fig:potential-kneg} marks the pivot positions.

We scan $k=-0.014,-0.015,-0.016,-0.017,-0.018$ and $N=50,55,60$.
Tables~\ref{tab:closedform-N50}--\ref{tab:closedform-N60} list all 15 points, and Table~\ref{tab:closedform-central} collects the central benchmark. The ranges are
\begin{align}
0.96473171&\le n_s\le0.97257986,\\
1.17719\times10^{-3}&\le r\le1.73525\times10^{-3},\\
5.14758\times10^{-6}&\le\alpha_s\le1.06759\times10^{-5}.
\end{align}
The field values range from $-1.00707305$ to $-0.69532998$, with $0.99299133\le V_*/M^4\le0.99569391$. Positive running here follows from $-2<x_*<0$ rather than from $k<0$ alone. The digits report the numerical evaluation of leading slow-roll expressions; they do not imply the same accuracy for the exact spectra.

\begin{table}[!ht]
\centering
\caption{Slow-roll observables for $N=50$ and the same prescribed endpoint $x_e=-2$. Digits are shown for reproducibility, not physical precision.}
\label{tab:closedform-N50}
\begin{tabular}{rccc}
\toprule
$k$ & $r$ & $n_s$ & $\alpha_s$\\
\midrule
-0.014 & $1.5463\times10^{-3}$ & 0.97258 & $5.1888\times10^{-6}$\\
-0.015 & $1.6108\times10^{-3}$ & 0.97064 & $6.3645\times10^{-6}$\\
-0.016 & $1.6634\times10^{-3}$ & 0.96871 & $7.6730\times10^{-6}$\\
-0.017 & $1.7047\times10^{-3}$ & 0.96677 & $9.1115\times10^{-6}$\\
-0.018 & $1.7353\times10^{-3}$ & 0.96484 & $1.0676\times10^{-5}$\\
\bottomrule
\end{tabular}
\end{table}

\begin{table}[!ht]
\centering
\caption{Slow-roll observables for $N=55$ and the same prescribed endpoint $x_e=-2$. Digits are shown for reproducibility, not physical precision.}
\label{tab:closedform-N55}
\begin{tabular}{rccc}
\toprule
$k$ & $r$ & $n_s$ & $\alpha_s$\\
\midrule
-0.014 & $1.3494\times10^{-3}$ & 0.97255 & $5.1896\times10^{-6}$\\
-0.015 & $1.3921\times10^{-3}$ & 0.97061 & $6.3250\times10^{-6}$\\
-0.016 & $1.4237\times10^{-3}$ & 0.96867 & $7.5771\times10^{-6}$\\
-0.017 & $1.4450\times10^{-3}$ & 0.96673 & $8.9409\times10^{-6}$\\
-0.018 & $1.4567\times10^{-3}$ & 0.96478 & $1.0410\times10^{-5}$\\
\bottomrule
\end{tabular}
\end{table}

\begin{table}[!ht]
\centering
\caption{Slow-roll observables for $N=60$ and the same prescribed endpoint $x_e=-2$. Digits are shown for reproducibility, not physical precision.}
\label{tab:closedform-N60}
\begin{tabular}{rccc}
\toprule
$k$ & $r$ & $n_s$ & $\alpha_s$\\
\midrule
-0.014 & $1.1772\times10^{-3}$ & 0.97253 & $5.1476\times10^{-6}$\\
-0.015 & $1.2027\times10^{-3}$ & 0.97058 & $6.2339\times10^{-6}$\\
-0.016 & $1.2182\times10^{-3}$ & 0.96863 & $7.4208\times10^{-6}$\\
-0.017 & $1.2244\times10^{-3}$ & 0.96668 & $8.7013\times10^{-6}$\\
-0.018 & $1.2224\times10^{-3}$ & 0.96473 & $1.0068\times10^{-5}$\\
\bottomrule
\end{tabular}
\end{table}

\begin{table}[!ht]
\centering
\caption{Dependence on the remaining slow-roll e-fold count at $k=-0.016$ and fixed $x_e=-2$.}
\label{tab:closedform-central}
\begin{tabular}{rccc}
\toprule
$N$ & $r$ & $n_s$ & $\alpha_s$\\
\midrule
50 & $1.6634\times10^{-3}$ & 0.96871 & $7.6730\times10^{-6}$\\
55 & $1.4237\times10^{-3}$ & 0.96867 & $7.5771\times10^{-6}$\\
60 & $1.2182\times10^{-3}$ & 0.96863 & $7.4208\times10^{-6}$\\
\bottomrule
\end{tabular}
\end{table}

\begin{figure*}[htbp]
\centering
\includegraphics[width=0.90\textwidth]{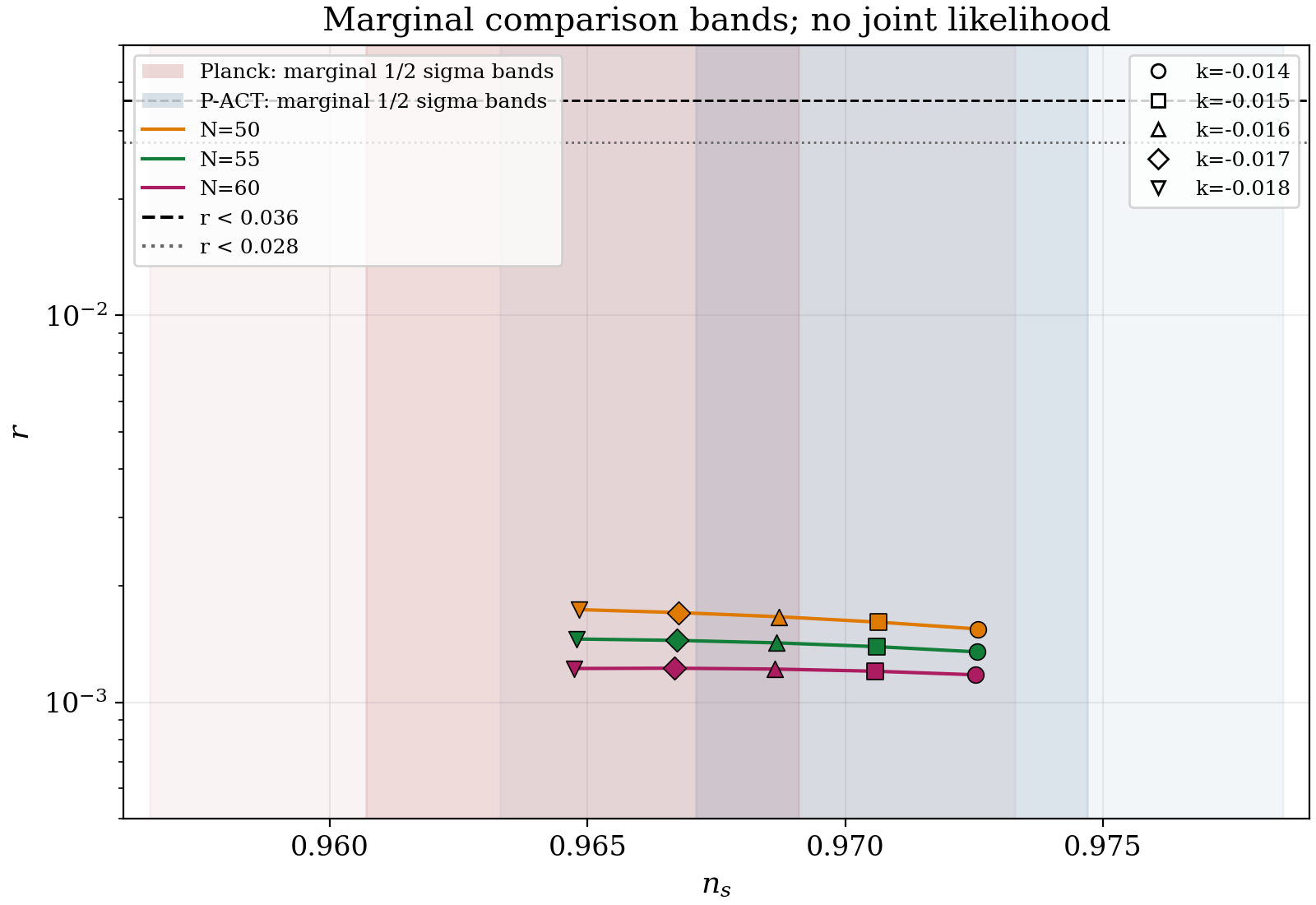}
\caption{Scan in the $(n_s,r)$ plane for fixed $x_e=-2$. Colors identify $N$ and markers identify $k$. Vertical shading shows the published marginal $1\sigma$ and $2\sigma$ intervals for Planck 2018 and P-ACT, solely as comparison bands; these are not joint posterior contours. The horizontal lines show the quoted $95\%$ tensor bounds $r<0.036$ and $r<0.028$. Each marginal $1\sigma$ interval contains 9 points, and each marginal $2\sigma$ interval contains all 15.}
\label{fig:ns-r}
\end{figure*}

Figure~\ref{fig:ns-r} compares the grid with published one-dimensional marginalized intervals. The Planck interval $[0.9607,0.9691]$ and the P-ACT interval $[0.9671,0.9747]$ each contain 9 of the 15 points at $1\sigma$; the respective quoted $2\sigma$ intervals contain all 15. This comparison does not determine joint $(n_s,r)$ credibility. The tensor ratios are below $0.036$ by factors of approximately $21$--$31$, rather than two orders of magnitude. We use no reconstructed Gaussian posterior or public CMB chains.

\begin{figure*}[htbp]
\centering
\includegraphics[width=0.80\textwidth]{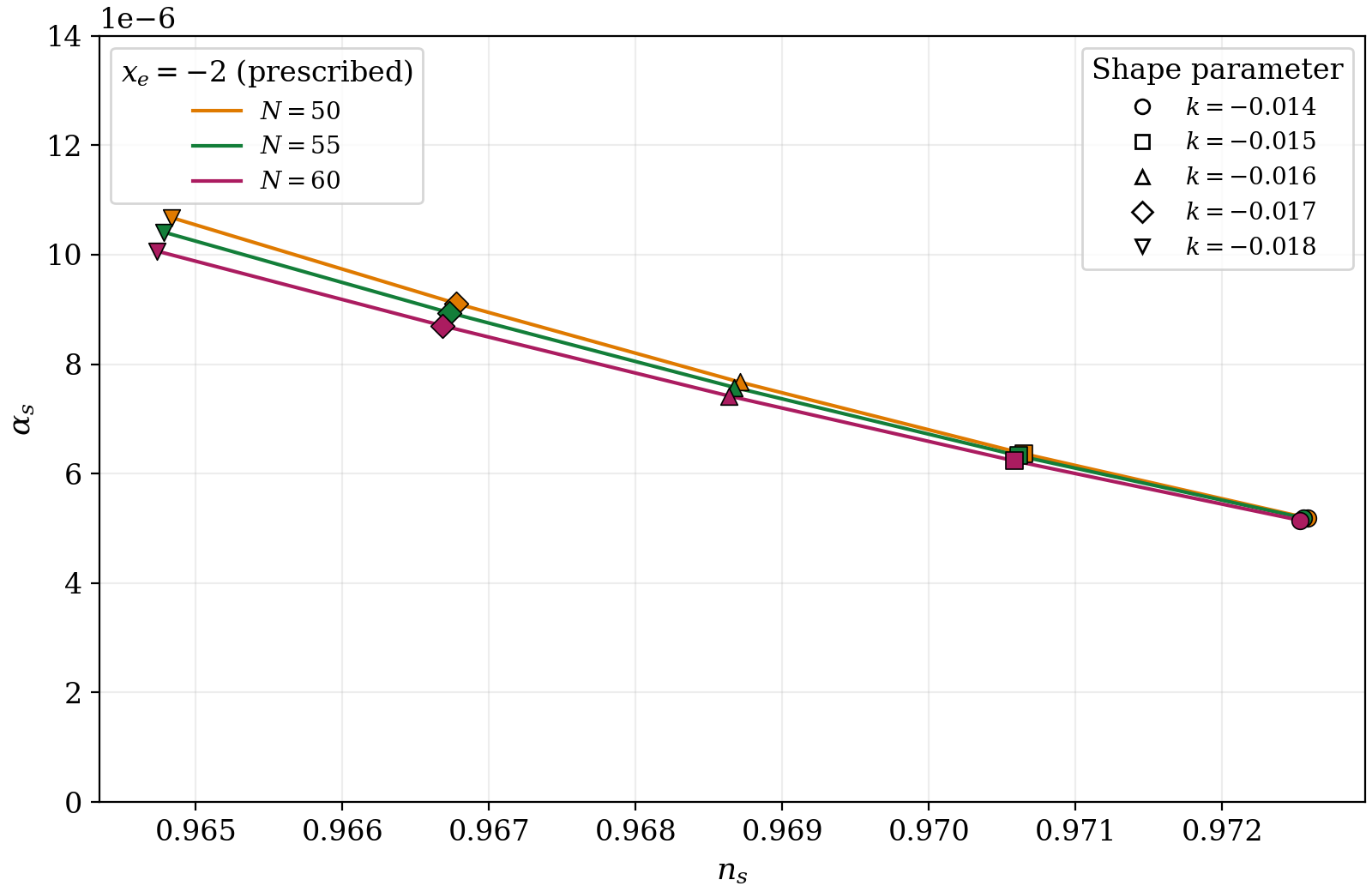}
\caption{Scalar running versus tilt for the fixed-endpoint grid. Colors identify $N$ and markers identify $k$, as in Fig.~\ref{fig:ns-r}. The running is positive because all sampled pivots satisfy $-2<x_*<0$, and is much smaller than the quoted current uncertainty scale. Positive running does not extend to the full negative-field branch.}
\label{fig:ns-alphas}
\end{figure*}

Figure~\ref{fig:ns-alphas} displays the small positive running in the sampled window. Its magnitude is well below the quoted Planck uncertainty scale $6.7\times10^{-3}$. This comparison checks consistency. It provides no evidence for detecting running or for a scale-invariant scalar spectrum, since the tilt itself remains different from unity.

\begin{figure*}[htbp]
\centering
\includegraphics[width=0.80\textwidth]{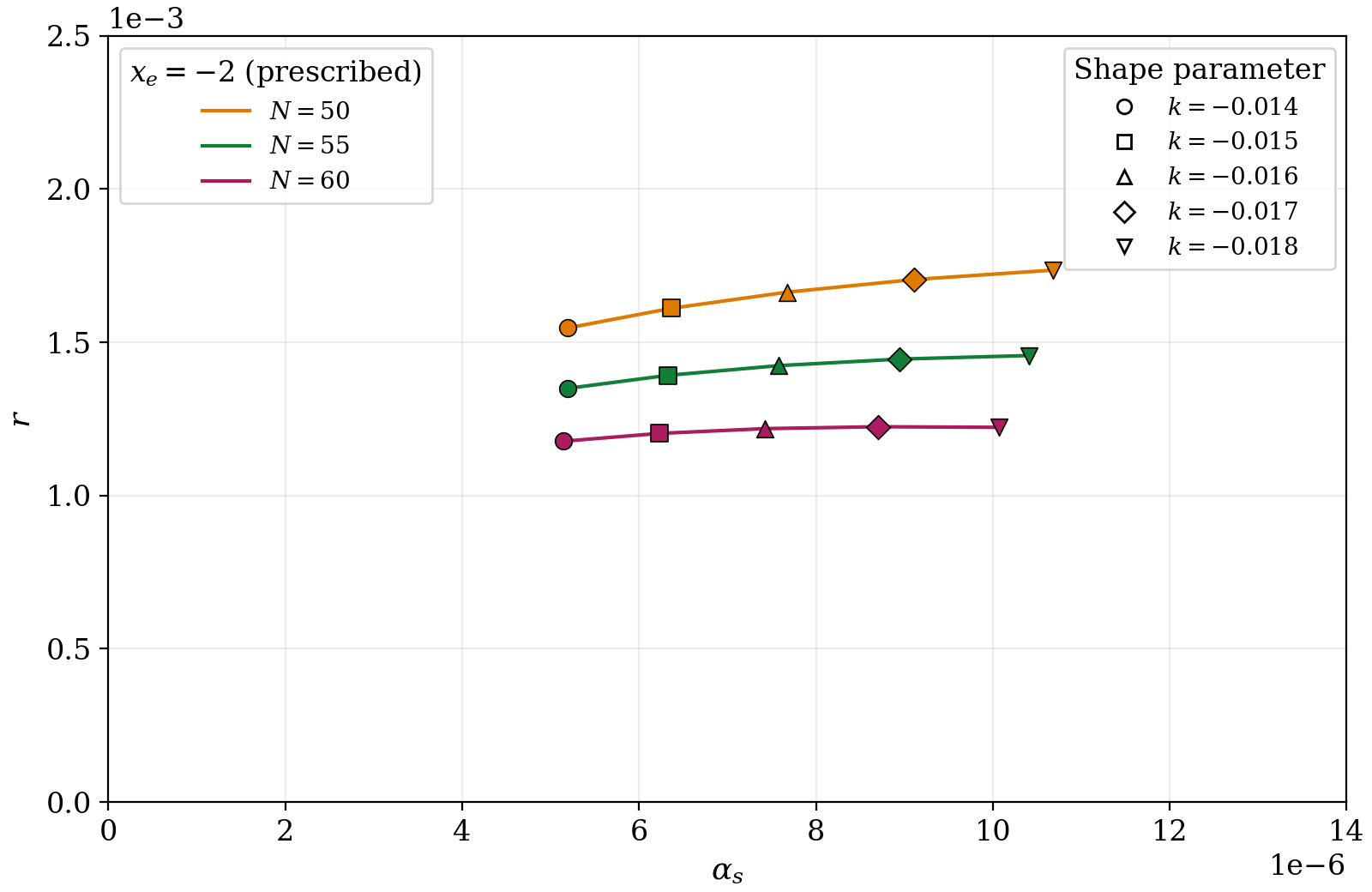}
\caption{Tensor ratio versus scalar running for the grid at fixed $x_e=-2$. Colors identify $N$ and markers identify $k$. The points are projections of the same local consistency relation; no universal monotonic increase of both observables with $N$ or $|k|$ is assumed.}
\label{fig:r-alphas}
\end{figure*}

Figure~\ref{fig:r-alphas} is a further projection of the same grid, constrained by Eq.~\eqref{eq:local-consistency}. At fixed $k$, $r$ decreases as the remaining slow-roll $N$ increases. Across this grid, neither the running as a function of $N$ nor $r$ as a function of $|k|$ follows a single strict monotonic trend.

The reheating diagnostic uses $k=-0.016$, $N_*=55$, $A_s=2.1\times10^{-9}$, $M_{\rm Pl}=2.435\times10^{18}\,\GeV$, $a_0=1$, $q_*=0.05\,\Mpc^{-1}$, $T_0=2.7255\,{\rm K}=2.34865418\times10^{-13}\,\GeV$, and $g_{s0}=3.909$. We hold $g_*=g_{s,\rm reh}=106.75$ throughout the illustration; this is not a realistic low-temperature thermal history. The conversions are $1\,\Mpc=3.0856775814913673\times10^{22}\,{\rm m}$ and $\hbar c=1.973269804\times10^{-16}\,\GeV\,{\rm m}$, giving $q_*=3.19746596\times10^{-40}\,\GeV$. Equations~\eqref{eq:Hstar} and~\eqref{eq:Vstar} then give $H_*=9.35302784\times10^{12}\,\GeV$ and $V_*/M_{\rm Pl}^4=4.42616681\times10^{-11}$. These inputs specify every curve through Eq.~\eqref{eq:rho-match}, without an imported additive constant.

\begin{figure*}[htbp]
\centering
\includegraphics[width=\textwidth]{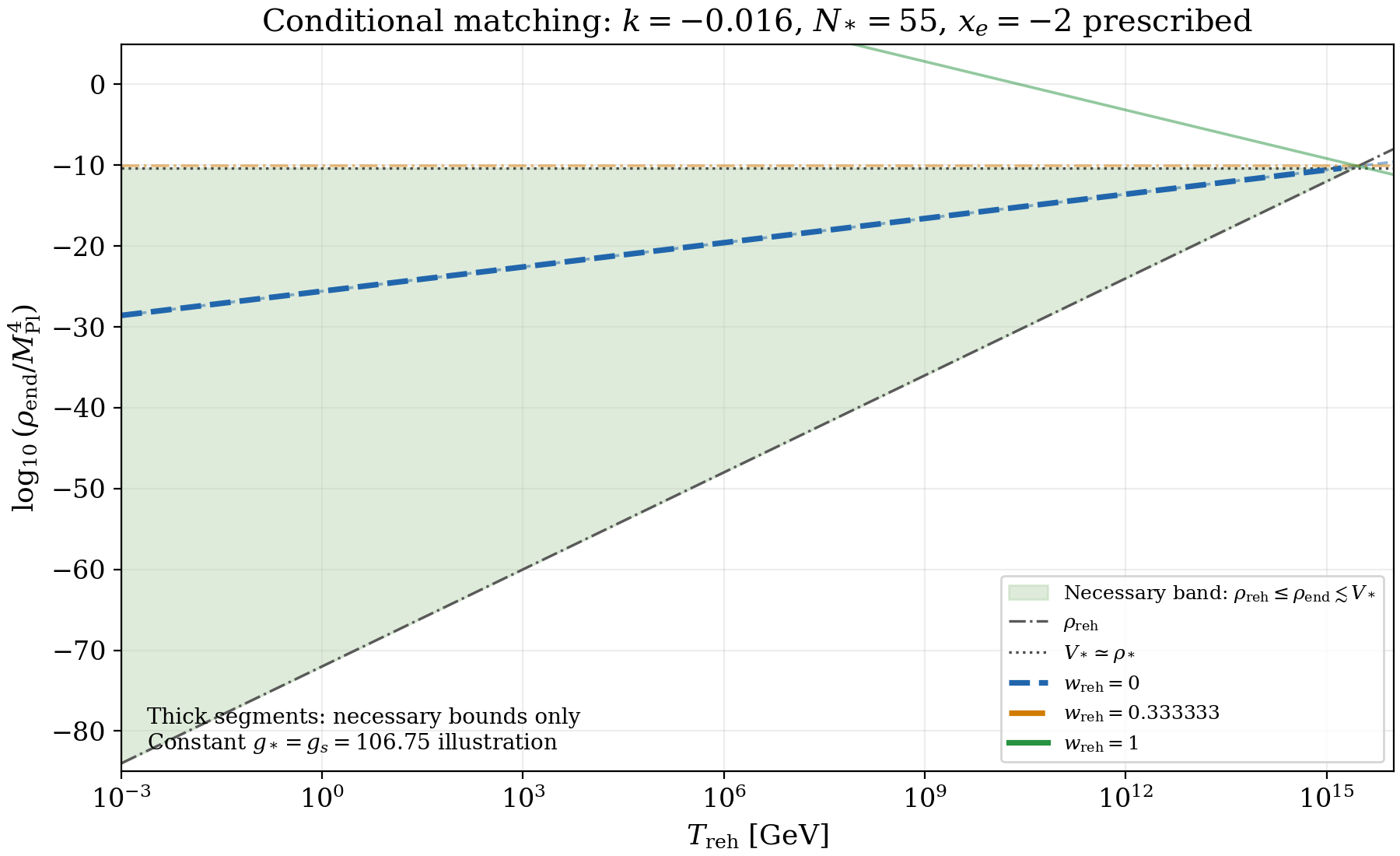}
\caption{Conditional reheating matching for the stated $k=-0.016$, $N_*=55$ benchmark and constant thermal degrees of freedom. The curves show the endpoint density inferred from Eq.~\eqref{eq:rho-match} for $w_{\rm reh}=0,1/3,1$, with slopes $+1,0,-2$ in $\ln\rho_{\rm end}$ versus $\ln T_{\rm reh}$. The band imposes both $\rho_{\rm reh}\le\rho_{\rm end}$ and $\rho_{\rm end}\lesssim V_*\simeq\rho_*$. Thick segments satisfy these necessary bounds and $N_{\rm reh}\ge0$; thin segments fail at least one condition. The plotted temperature range starts at $1\,\MeV$ as a reference, not a sufficient BBN criterion. No complete exit model is implied.}
\label{fig:reheating-param}
\end{figure*}

With these illustrative conventions, the matter-like curve has a segment satisfying the necessary bounds, whereas the radiation-like and kinetic curves do not. These matching conditions do not select kinetic-dominated reheating or exclude other values in a continuous $w_{\rm reh}$ range. The freely inferred $\rho_{\rm end}^{\rm match}$ has not, however, been connected to the total energy of a completion at the imposed field endpoint $x_e=-2$. The figure therefore cannot establish which reheating histories the model actually realizes.

\subsection{Finite exact-background checks and outstanding spectral tests}

The accompanying script integrates Eq.~\eqref{eq:exact-phase} for $k=-0.018,-0.016,-0.014$ and each $N=50,55,60$, after a slow-roll initialization and a 15-e-fold pre-roll estimate before the pivot. Repeating with a 25-e-fold pre-roll gives the same pivot velocities and endpoint durations to the reported numerical tolerance. The integrator uses relative tolerance $2\times10^{-11}$, absolute tolerance $2\times10^{-13}$, and maximum step $0.1$ in $n=\ln a$. Over these nine cases, the exact duration from the tabulated $x_*$ to $x_e=-2$ exceeds the slow-roll label by $0.22322$--$0.34269$ e-folds. At $k=-0.016$, the exact durations are $50.25401$, $55.27981$, and $60.30567$. The endpoint still has $\epsilon_H<5.98\times10^{-4}$ and is therefore not a dynamical end of inflation. A bare-model continuation of the central pivot for 1000 e-folds remains inflationary, consistently with the analytic invariant-strip proof; this finite test is not the proof itself.

These background checks quantify the e-fold error over the stated sample. They do not determine a second-order or exact-mode error for $(n_s,r,\alpha_s)$, or the transfer of perturbations through an unspecified exit. Those calculations, together with a full comparison with the closest related potential, remain necessary before claiming precision predictions, priority, or a complete cosmological realization.

\FloatBarrier
\section{Comparison with representative inflationary classes}
\label{sec:comparison}

The present construction should be compared with benchmark potentials, not because any one model is assumed to be preferred, but because the comparison clarifies which features of the predictions arise from the potential geometry.

For a monomial potential
\begin{equation}
V\propto\phi^p,
\end{equation}
the slow-roll parameters are typically controlled by inverse powers of the field and the tensor amplitude can be comparatively large.

For an exponential potential
\begin{equation}
V\propto e^{-\lambda\phi/M_{\rm Pl}},
\end{equation}
one has constant
\begin{equation}
\epsilon_V=\frac{\lambda^2}{2},
\end{equation}
so standard slow-roll inflation requires $\lambda^2\ll1$ and does not naturally end without an additional mechanism.

For plateau models such as Starobinsky inflation,
\begin{equation}
V(\phi)
=
V_0
\left(
1-e^{-\sqrt{2/3}\phi/M_{\rm Pl}}
\right)^2,
\end{equation}
the observable regime occurs at large field values with a parametrically small slope \cite{Starobinsky1980}.

The exponential power-law model has a finite-field hilltop and a nonzero extremum. Its local structure is summarized by Eq.~\eqref{eq:local-consistency}, while the negative-field exit obstruction follows from Eq.~\eqref{eq:exact-phase}. These properties distinguish its description from the pure exponential and the displayed Starobinsky potential, but do not establish uniqueness among generalized exponential or hilltop families. A full comparison with Ref.~\cite{Younesizadeh2026GRG} requires its explicit potential, independent shape parameters, branch, endpoint, and observable relations to be examined. The accessible abstract and bibliographic record alone do not settle this comparison and provide no basis for a priority claim.

\FloatBarrier
\section{Discussion}
\label{sec:discussion}

The model has several features that simplify its analytic treatment. First, its logarithmic derivatives are elementary, which permits a largely analytic treatment. Second, the exponent parameter $k$ continuously controls the departure from the vacuum-energy limit. Third, the normalization of the scalar spectrum can be imposed algebraically once the inflationary trajectory is specified.

At the same time, the analysis reveals a structural limitation of the potential. For $k<0$ the stationary point at $\phi=0$ is a maximum rather than a minimum, the field rolls away from it, and the potential has no local minimum at all. The potential therefore does not by itself supply a complete inflationary plus reheating history.

This conclusion also changes the interpretation of a naive scan based solely on $n_s$ and $r$. A point in the observationally interesting region is not sufficient to establish model viability. The same point must have a consistent field trajectory, a specified end of inflation, a controlled energy density at the end of inflation, and a reheating history that connects to the standard hot Big Bang.

The parameterization in Sec.~\ref{sec:closedform-scan} fixes one phenomenological endpoint across the grid and uses the physical slow-roll e-fold integral. That section reports the numerical ranges and interval counts; the schematic comparison regions are not interpreted as measured joint posteriors. Positive running holds in the sampled window, not on the whole negative-field branch. Any proposed discrimination based on its small magnitude must account for truncation error and observational sensitivity.

As $x\to-\infty$, $\epsilon_V\to1/2$ and $\eta_V\to1$, outside controlled slow roll. Substitution into first-order formulas formally gives $r\to8$ and $n_s\to0$, but these are not controlled exact spectral predictions. This tail is outside the observational slow-roll window. The exact no-exit proposition explains why an additional sector is required; the reheating formulas for an oscillating quadratic minimum apply only to a specified completion.

Scale matching does not establish that the exit must be kinetic-dominated. The three reference equations of state in Fig.~\ref{fig:reheating-param} are diagnostics within a continuous parameter space. Both energy bounds, nonnegative duration, and consistency with the energy of the chosen endpoint must hold. Satisfying these necessary conditions neither specifies the exit sector nor establishes preservation of the primordial adiabatic spectrum.

The reheating temperature in a fully specified completion would be obtained from a pivot-matching formula and a constant-$\wre$ parameterization. It is not predicted by the potential alone. A complete particle-physics realization must specify the decay channels and verify that the assumed effective equation of state and thermalization history are realized.

For this reason, the physically meaningful parameter space is more naturally represented as
\begin{equation}
\mathcal P_{\rm physical}
=
\left\{
(k,x_e,N_*,\wre,T_{\rm reh},M,\text{exit inputs})
\;|\;
A_s,n_s,r,\rho_{\rm end},\rho_{\rm reh}
\ \text{are mutually consistent}
\right\}.
\end{equation}
The precise allowed region should be obtained numerically for a specified completion.

\FloatBarrier
\section{Conclusions}
\label{sec:conclusions}

We have presented a systematic analysis of canonical single-field inflation based on the exponential power-law potential
\begin{equation}
V(\phi)=M^4
\exp\!\left(\frac{\phi}{M_{\rm Pl}}\right)
\left(
1+k\frac{\phi}{M_{\rm Pl}}
\right)^{-1/k},
\end{equation}
with $k<0$. The construction combines exponential and power-law structures and is sufficiently simple to permit closed expressions for the main inflationary quantities.

Introducing $x=\phi/M_{\rm Pl}$, we find
\begin{equation}
\epsilon_V
=
\frac12
\left(
\frac{kx}{1+kx}
\right)^2,
\end{equation}
and
\begin{equation}
\eta_V
=
\frac{k}{(1+kx)^2}
+
\left(
\frac{kx}{1+kx}
\right)^2.
\end{equation}
The scalar spectral index and tensor-to-scalar ratio follow as
\begin{equation}
n_s-1
=
\frac{2k}{(1+kx_*)^2}
-
\left(
\frac{kx_*}{1+kx_*}
\right)^2,
\end{equation}
and
\begin{equation}
r
=
8
\left(
\frac{kx_*}{1+kx_*}
\right)^2.
\end{equation}
The scalar amplitude fixes the scale $M$ once the horizon-exit point is determined.

The field-space structure is equally relevant. The potential has a stationary point at $\phi=0$ with $V(0)=M^4$. For $k<0$ this is a maximum and the field is driven away from it. The real-domain boundary at $\phi=-M_{\rm Pl}/k$ lies on the positive-field side, where the potential tends to zero; it is generally a complex branch point, not a divergent pole of $V$. Consequently, the potential does not automatically produce the standard sequence from slow-roll inflation to oscillatory reheating and radiation domination. A consistent cosmological realization must specify an exit mechanism.

The analysis gives a local relation among $(n_s,r,\alpha_s)$ at specified slow-roll orders, an exact no-exit statement for downhill inflationary initial data, and a Lambert-$W$ parameterization with an explicit endpoint. Tables~\ref{tab:closedform-N50}--\ref{tab:closedform-central} and Figs.~\ref{fig:potential-kneg}--\ref{fig:reheating-param} give conditional examples at fixed $x_e=-2$; their numerical ranges are collected in the results section. The endpoint serves as a matching input; no mechanism implementing it has been constructed. The asymptotic exponential tail is outside the controlled observational slow-roll regime.

Explicit conventions make the reheating matching reproducible. Its necessary physical bounds include $\rho_{\rm reh}\le\rho_{\rm end}\le\rho_*$ and $N_{\rm reh}\ge0$; it does not select a unique effective equation of state. A complete realization still requires an exit sector that connects the endpoint energy to the thermal history, checks adiabatic evolution of the observable modes, and quantifies higher-order and exact-mode corrections. Claims of complete cosmological viability or priority over the closest generalized exponential models are therefore deferred.

\FloatBarrier
\appendix

\section{Useful exact identities}
\label{app:identities}

For
\begin{equation}
V=M^4e^x\Delta^{-1/k},
\qquad
\Delta=1+kx,
\end{equation}
the following identities are useful:
\begin{align}
\frac{V_{,\phi}}V
&=
\frac{1}{M_{\rm Pl}}
\frac{\Delta-1}{\Delta},\\
\frac{V_{,\phi\phi}}V
&=
\frac{1}{M_{\rm Pl}^2}
\left[
\frac{k}{\Delta^2}
+
\frac{(\Delta-1)^2}{\Delta^2}
\right],\\
\epsilon_V
&=
\frac{(\Delta-1)^2}{2\Delta^2},\\
\eta_V
&=
\frac{k+(\Delta-1)^2}{\Delta^2}.
\end{align}

The number of e-folds is
\begin{equation}
N
=
\frac1k
\left[
\Delta_*-\Delta_e
+
\ln\left|
\frac{\Delta_*-1}{\Delta_e-1}
\right|
\right].
\end{equation}

\section{Local expansion around the stationary point}
\label{app:minimum}

Expanding the logarithm,
\begin{align}
\ln\frac{V}{M^4}
&=
x-\frac1k\ln(1+kx)\\
&=
\frac{kx^2}{2}
-\frac{k^2x^3}{3}
+\frac{k^3x^4}{4}
+\mathcal{O}(x^5).
\end{align}
Consequently,
\begin{align}
\frac{V}{M^4}
&=
1+\frac{kx^2}{2}
-\frac{k^2x^3}{3}
+\left(\frac{k^3}{4}+\frac{k^2}{8}\right)x^4
+\mathcal{O}(x^5).
\end{align}
For $k<0$ the stationary point is a quadratic maximum, and the field is unstable at $x=0$.

\section{Observational normalization}
\label{app:normalization}

Given $(k,x_*)$, the amplitude condition gives
\begin{equation}
M
=
M_{\rm Pl}
\left[
24\pi^2A_s\epsilon_{V*}
e^{-x_*}
(1+kx_*)^{1/k}
\right]^{1/4}.
\end{equation}
Then
\begin{equation}
V_*
=
M^4e^{x_*}(1+kx_*)^{-1/k},
\end{equation}
and
\begin{equation}
H_*
=
\sqrt{\frac{V_*}{3M_{\rm Pl}^2}}.
\end{equation}

\section{Consistency checks for a numerical scan}
\label{app:scan}

For each numerical point, the following conditions should be checked:
\begin{align}
1+kx_*&>0,\\
0&<\epsilon_{V*}<1,\\
|n_s-0.9649|&\lesssim\text{chosen observational tolerance},\\
r&<r_{\rm max},\\
A_s&\simeq2.1\times10^{-9},\\
T_{\rm reh}&\gg \mathcal O(\MeV),\\
\rho_{\rm reh}&\le\rho_{\rm end}\le\rho_*,\\
N_{\rm reh}&\ge0,
\end{align}
where the final numerical tensor and scalar constraints should be updated to the exact dataset and pivot scale used in the final analysis. Every point must use the same specified $x_e$ and exit prescription. A completion must verify that the exit preserves the relevant adiabatic modes and the superhorizon spectrum. These checks are necessary but not sufficient for a full likelihood analysis.

\section{Scope and limitations}
\label{app:scope}

The numerical results concern local slow-roll spectra and a fixed phenomenological endpoint, together with necessary thermal matching conditions. They do not constitute a completed cosmological solution. Exact background integration tests the no-exit proposition and the e-fold truncation error over the stated finite sample, but is not an exact perturbation calculation. No exact scalar or tensor mode integration, second-order spectral error budget, or exit transfer calculation is supplied here. Precision use of the consistency relation or a claim of cosmological viability still requires those checks. In particular, choosing $x_e=-2$ does not end inflation in the bare potential, and the inferred matching density in Fig.~\ref{fig:reheating-param} must still be reproduced by an explicit completion. Full-text comparison with the closest related potential is also required before a priority or uniqueness claim.

\section{Complex structure of the slow-roll differential forms}
\label{app:complex}

On a chosen branch of the complexified field $z$, the logarithmic derivative and e-fold differential are
\begin{align}
\dd\ln V&=\left(1-\frac{1}{1+kz}\right)\dd z,
&\mathop{\rm Res}_{z=-1/k}\dd\ln V&=-\frac1k,\\
\dd N&=\left(1+\frac{1}{kz}\right)\dd z,
&\mathop{\rm Res}_{z=0}\dd N&=\frac1k.
\end{align}
The logarithm in Eq.~\eqref{eq:Nx} and its Lambert-$W$ inverse follow from the second expression. Analytic continuation of $V$ once around $z=-1/k$ multiplies it by $e^{-2\pi i/k}$. When $p=-1/k$ is a positive integer, this point is an analytic zero of order $p$; otherwise the potential is generally multivalued. The argument principle counts integer zeros minus poles only for a suitable single-valued meromorphic function and contour. An arbitrary real $k$ therefore does not imply quantized winding of $V$, nor do these formal residues alone define a physical entropy or temperature.


\vspace{1em}
\noindent\textbf{Data Availability Statement.}
The accompanying source package contains the calculation and plotting scripts, numerical tables, parameter conventions, and finite background checks used for the figures. The observational comparison uses published marginalized constraints; no public CMB posterior chains or likelihood data are analyzed. The comparison bands are not joint credible regions.

\vspace{1em}
\noindent\textbf{Funding.}
This research received no specific grant from any funding agency in the
public, commercial, or not-for-profit sectors.

\vspace{1em}
\noindent\textbf{Conflict of Interest.}
The authors declare no competing interests.


\end{document}